\documentclass[aps,prl,twocolumn,showpacs,superscriptaddress,groupedaddress]{revtex4}
\usepackage{graphics}
\usepackage[pdftex]{graphicx}
\usepackage[pdftex]{epsfig}
\usepackage{epstopdf}
\usepackage{bm}
\usepackage{color}
\usepackage[capitalize]{cleveref}
\begin{document}
%\linenumbers

\title{Comparative analysis of methods to estimate the tire/road friction coefficient applied to traffic accident reconstruction}

\author{A. Baena}

\affiliation{Departamento de F\'isica,  Universidad Antonio Nariño, Bogot\'a, Colombia}

\author{E. Remolina}

\affiliation{Centro de Formaci\'on en Tr\'ansito y Transporte CIFTT, Bogot\'a, Colombia}

\author{H. Londoño}

\affiliation{Departamento de F\'isica,  Universidad Antonio Nariño, Bogot\'a, Colombia}

\author{G. Enciso}

\affiliation{}

\author{W. Toresan}

\affiliation{Centro de Entrenamiento en Investigaci\'on y Reconstrucci\' on de Accidentes, Resistencia, Argentina}

\date{\today}
%\affiliation{Centro de Entrenamiento en Investigaci\'on y Reconstrucci\' on de Accidentes, Resistencia CEIRAT, Argentina}

\begin{abstract}
The measure of the friction factor in the location of the event is very important to get accurate inputs to model the accident, as the friction factor depends on environmental conditions and the dynamics of the event. In some low-middle income countries, accident reconstruction experts, usually don’t have the feasibility to acquire expensive equipment as high precision accelerometers; therefore, it is important to identify a low-cost method that provides data quality comparable with high precision equipment. 
This study presents a comparison of three methods: VC4000PC accelerometer, sensor kinetics mobile app and video analysis by tracker (free software). The methods have been compared experimentally for emergency braking with blocked wheel, as well as using ABS. Data from the three methods were recorded simultaneously. Data analysis was made applying inferential statistics. The results indicate that the data collected with the smartphone app ensure good accuracy and does not provide significant variance in comparison with the accelerometer. On the other hand, the data obtained by video analysis shows a good linear relationship with the accelerometer; however, it shows some statistical evidence of differences related to precision and variances in comparison with accelerometer device. The confidence interval, absolute error, and other estimators have been obtained for the smartphone method, to promote the optimal using by the experts to get parameters on site and apply to a traffic accident reconstruction.

\end{abstract}
\maketitle
\pagebreak

\section{Introduction}
During a traffic accident investigation, it is necessary using some input parameters which leave to reconstruct the fact closer to reality in hindsight to know the causes of the accident. The use of the appropriate friction coefficient or drag factor which determines the interaction between tire/road is crucial for modeling, considering that it depends on different context conditions, such as road, environmental and the dynamics of each event. Factors as moisture and type of surface of the road affect the distance of the tire sliding during the braking maneuver rising vertical curve, horizontal curve with cant and many other situations. Those factors are integrated into the input friction coefficient \cite{Rivers2}.

Since the '70s have been reported many methods to measure the coefficient of friction proposed by the traffic accident reconstruction community.  Many of them have been established by the ASTM International (American Society for Testing and Materials).  For instance, the \textit{Skid Tester}, which consists of a tire pulled by a vehicle \cite{skidtester}. The \textit{Mu-Meter}, that be found in a semi-trailer and a central tire pulled by a vehicle (ASTM E670) \cite{mumeter}. Similar working principle have the \textit{Skid Trailer} (ASTM 274) \cite{skidtrailer}. The \textit{Sand Patch Test} (ASTM E965), being two small tires pushed by a manual force applied through a bar and the \textit{3D laser scanning} utilized to quantify the mean profile depth of pavement at a specific location \cite{sand}.  Another methods as \textit{Dynamic Friction Tester} (ASTM 1911) and \textit{British Pendulum} (ASTM E303) are widely used  to measure the texture of the surface. The first one consists of a horizontal spinning disk fitted with three spring-loaded rubber sliders which contact the paved surface and the second one measures skid resistance when a rubber slider on a  pendulum arm contacts a test surface. Previous works compare the mentioned methods as a function of the type of surface and environmental conditions \cite{friction}. The results showed a friction value in a good agreement between some of them, for instance, the  \textit{Dynamic Friction Tester} and the \textit{British Pendulum}.

As described above, despite the larger number of methods to estimate physical parameters as friction coefficient, the challenge to reconstruct a traffic accident is the design of a test recreating the same conditions of the accident under investigation. The ideal test to be used in a traffic accident reconstruction consist of the same vehicle, at the same spot, with the same load and surface conditions \cite{Fricke}. Although this optimal test design is especially difficult, a braking test with vehicles to measure cinematic parameters allows the control of some factors within the context conditions.

Accelerometers as \textit{Vericom}, \textit{XL Meter}, \textit{ g-analyst} have been widely recommended for this kind of test. In some pioneers works, mentioned accelerometers were compared establishing important precision differences related to the information quantity. As a result, the first versions of Vericom have been accepted as a complete sensor, which permits good accuracy measurements of the initial velocity, deceleration and time during a braking maneuver of a vehicle \cite{branch}. 

Despite that there are methods more appropriate than others for physical parameters estimation focusing on traffic reconstruction, the reality is that the method selected by the expert depends on factors like price, disruption of traffic, time and safety.  Many experts have not easy access to a high precision accelerometer. Therefore, they prefer to adopt the friction coefficient values from tables published in other contexts. The value is usually selected according to the types of vehicle and surface  \cite{Camiones},\cite{Rivers2},\cite{Fricke}. On another hand, the experts could obtain information about some parameters from the footprint of braking. However, nowadays, many vehicles have ABS brakes, in those cases, a trace of braking becomes difficult to observe.

Bearing the above considerations in mind, it emerged the interest to study different methods applying to the braking test, whose main features is to be more accessibility than accelerometers recommended in the protocols (e.g ASTM). One of the methods proposed was the use of a smartphone. Currently,  this device is one of the most used devices in the world for its portability, affordability and the high number of applications, such as the accelerometer sensors. Its application have been widely studied in different areas, such as experimental physics for teaching, medicine, geophysics, road safety, and sports \cite{Univalencia}, \cite{SM-Rebecca}, \cite{Bruwer},\cite{SMDETEC}.

Another method proposed in this work to be applied in braking tests with vehicles is through video processing. Currently, video capture and analysis have become a very feasible source of information. It has been used in many disciplines. For instance, free software as Tacker has been used in physics class experiments to measure cinematics variables focusing on modeling and studying the physics phenomena \cite{Unal,Navarrete, Ramli}. Tacker software to video analysis has even been applied in traffic reconstruction cases, for example, to measure the projection velocity of a pedestrian impacted by a vehicle, as well as to support the accident modeling and simulation \cite{Torres}.
This paper presents a comparative analysis of three methods, accelerometer VERICOM 4000PC, smartphone accelerometer Kinetics Sensor Pro and video processing by Tracker free software, used to measure deceleration and other variables during emergency braking tests with a vehicle. The measurements of tests are the inputs parameters to estimate the friction coefficient between tire/road to be applied to traffic accident reconstruction.

This paper is organized as follows: Section METHODOLOGY describes the procedures and protocols used to design the braking test with a vehicle; likewise, the features of methods, experimentation sites, and the vehicle. Section RESULTS presents the statistical models applied to the comparative analysis of methods, as well as the results. Section CONCLUSION AND DISCUSSION expounds the conclusions from the comparison between methods and the impact of the results on the traffic reconstruction community. Finally, Section APPENDIX, summarize the adjustment made to the procedures and protocols to include the other methods under this study.
\section{Methodology}
 
To the comparative analysis of the 3 methods (smartphone, video, and accelerometer) to measure the deceleration of a vehicle with its brakes fully applied, it has been made 8 experiments with approximately $10$ tests by each one at four different zones \cite{Dios12}. To design of experiments was considered the recommendations of the following protocols:
\newline
- ISO21994 for Stopping distance at straight-line braking with ABS -Open-loop test method \cite{ISO21994}.
\newline
- Measurement of Vehicle-Roadway Frictional Drag SAEJ2505 \cite{SAE2505}.
\newline
- Measurement of the deceleration factor in vehicle brakes with smartphone by CEIRAT.

Nevertheless, adjustments were set on the guideline of procedures, with the including the use of the smartphone and the video recording, as well as to adjust within the particular context. The code proposed is described with more details in the section Appendix.

\subsection{Deceleration Measurement Devices }

 \textit{Smartphone}: For this study were used iPhone 7 and iPhone X, with application Kinetics Pro Sensor to obtain the deceleration values. The among of data was 30 per second for the three axes. However, for the analysis, it has been used the longitudinal axis, namely, along the trajectory of the vehicle.
\newline

 \textit{Video Recording}: The video recording was made with iPhone 7 on a tripod by adjusting the focal plane. The camera characteristics are dual 12 MP with wide-angle and telephoto, recording video in 4K at 30 fps and HD 1080p at 30 cps with optical stabilization. Each video is imported into the free Tracker software to analyzes the deceleration of the vehicle. Data set obtained were 30 per second as the smartphone method.
\newline

 \textit{Accelerometer}:  Vericom VC4000PC Accelerometer in braking mode. The data was  100 per second and acquired by Profile 5 software. The device permits to obtain time, speed, distance and average deceleration of the vehicle.

\subsection{Braking Tests Conditions}
\subsubsection{Environmental Features}
The Experiments were made with the following context features:

\textit{Experiment 1}: Surface of concrete, bad conditions (bumps in the road), traffic worn and dry state.

\textit{Experiment 2}: Surface of asphalt, dry condition, and low traveled road.

\textit{Experiment 3}: Surface of asphalt and partly wet condition and loose material, traffic worn road.

\textit{Experiment 4 and 5}: Surface of asphalt, wet condition, rainy weather, and traffic worn road.

\textit{Experiment 6,7 and 8}: Surface of asphalt, dry condition, and low traveled road.

\begin{figure}[h]
	\centering
	\includegraphics[width=0.9\linewidth]{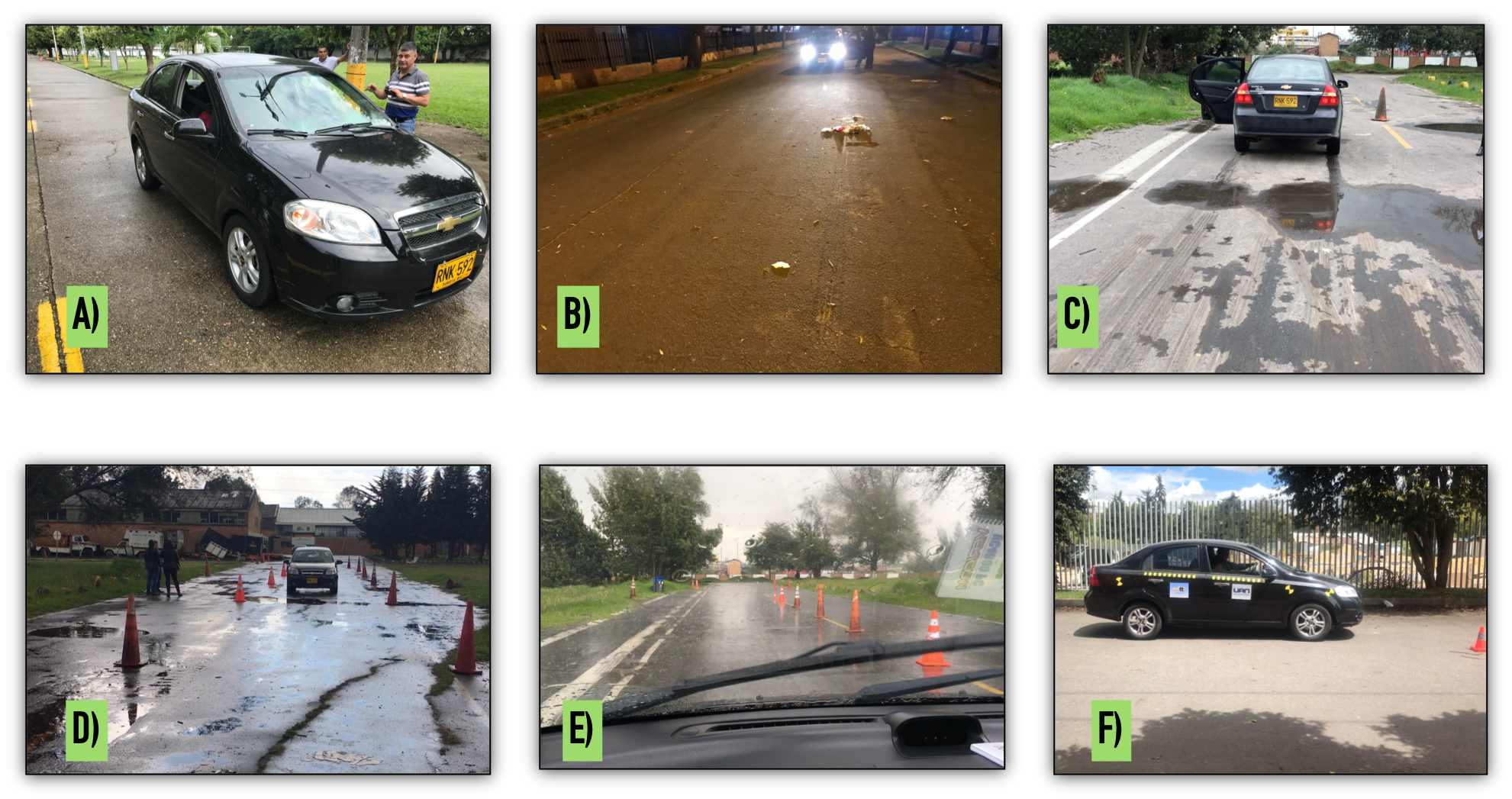}
	\caption{Context of Braking Tests. A) Experiment 1 ; B) Experiment 2; C) Experiment 3 D) Experiment 4; E) Experiment 5  and F) Experiments 6,7 and 8.}
	\label{fig:lExp-Context}
\end{figure}

\subsubsection{Vehicle and Speeds}
The vehicle tested was a Chevrolet Evolution 1.6, with hydraulic brake system with ABS and ventilated discs on front wheels and drum on rear tires, tires 185/55 R15; length 4310 mm, width 1450 mm, height 1495 mm, distance between axles 2480 mm, weight to vacuum or 1125 kg, load capacity 415 kg, full fuel tank 11,88/45 gal/L. General brand tires G-Max Rs 185/55 R15 82V (3-6), with inflation pressure at 32 PSI as specified by the technical code.
The accelerometer VC4000PC is able to record the speed at the beginning of the braking process. The speeds registered in the experiments are in the range of 25 Km/h and 45 Km/h.

%\begin{table}[h]
%\begin{tabular}{lclcl}
%Experiment 1 &   30 km/h  \\
% Experiment 2 &    \\
% Experiment 3 &    33,6 km/h \\
% Experiment 4 &   4
%	\label{tabla:velocidad}
%\end{tabular}
%\end{table}

\section{Results}
\noindent
In order to compare the methods using to measure the deceleration during the emergency braking tests, it has been made the same data treatment for each method. First of all, it was consider the deceleration as a function of time, starting from that the vehicle changes the velocity by the action of brakes until the vehicle is stopped. The curve was considered containing two sections. Each section presents a behavior related to the vehicular dynamic phenomena during the process.
\newline

\begin{figure}[h]
	\centering
	\includegraphics[width=1\linewidth]{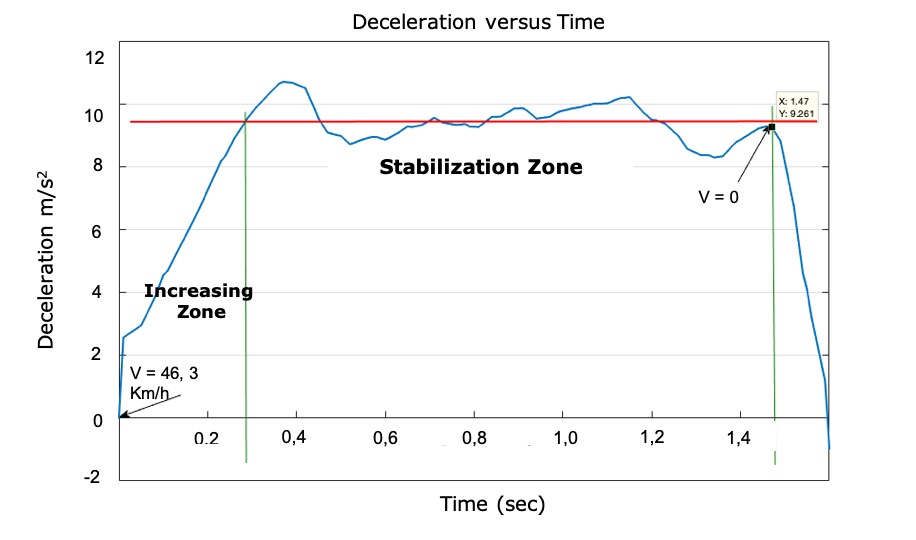}
	\caption{Deceleration as a function of time. Test 1 of experiment 8. Speed $46,3 km/h$. Data from AC VC4000PC. }
	\label{fig:grafic-p99}
\end{figure}

\noindent
Fig \ref{fig:grafic-p99} shows the deceleration ($m/s^2$) versus time ($s$) obtained by accelerometer in test 1 of the experiment 8. The begging to the end of the braking maneuver is observed. Two sections can be identified. The first section called here as \textbf{Increasing Zone IZ} corresponds to the deceleration (related to the friction coefficient) starting from 0 until a mean value of deceleration (red curve) at the reaction time of the driver (green curve). This zone represents the first moment when the driver applies the brake pedal force until the production of sufficient one to cause wheels with conventional brakes to lock and/or for the effective operation of the ABS.
\newline

\noindent
The second section called the  \textbf{Stabilization Zone SZ} is the flat region of the braking process. Despite there are fluctuations, it could assume all values at "flat zone" are fluctuating around an expected value of deceleration (red curve).   In many cases of traffic accidents,  it is possible to observe marks on the road surface or skidmarks (dark shadow), which corresponds to the initialization of the stabilization zone. The skidmarks will intensify during the sliding of locked wheels over the asphaltic surface.  It is also observed sometimes when the ABS has been activated, nonetheless, the marks are discontinuous according to the periodic locking of ABS. In all cases, the marks may have grooves shapes depending on the characteristics of the road

The first step of the analysis is extracting the SZ and estimating the moving average, as described following.

\subsection{Estimation of Moving Average for Stabilization Zone SZ}
To extract the SZ, data were treatment manually, considering as the beginning of the SZ the point of the mean value where the fluctuations are concentrated (red line Fig \ref{fig:grafic-p99}). In this zone, it could assume the deceleration fluctuating around a value. To obtain that mean value, the moving average has been calculated. The treatment is a prediction technique where a series of averages are created, each of them taking into account the previous data. The treatment smoothes the fluctuations of a short-term time series, highlighting trends or long-term cycles. This estimation was applied for all 89 braking tests with the three methods corresponding to 267 data sets. 
The statistical model is defined as the average into a period time as in Eq. \ref{fig:MV}

\begin{equation}
 \widehat{ X}_i = \frac{\sum_{i=2}^{n} X_{i-1}} {n}
\label{fig:MV}
\end{equation}

$\widehat{X}_i$ is the period of desaceleration at $i=2$; $X_{i-1}$ previous periods at $i$, and $n$ is the number of data.

The mean value of deceleration has been obtained for each method (accelerometer, smartphone, and video analysis) aand for each experiment. The values were compared through its statistical parameters \cite{Dios12}, such as variance, standard error, standard deviation, minimum value, maximum value, assuming a level of confidence of 95\% \cite{Dios10}.  Table \ref{tabla:compromedio2} shows the statistical parameters obtained for method smartphone, experiment 8. Table, \ref{tabla:compromedio4}, presents the global statistical parameters of the three methods calculated for experiment 8.

\begin{table}[htbp]
	\begin{center}
		\resizebox{9cm}{!} {
			\begin{tabular}{|l|c|c|c|c|c|c|c|c|c|c|}
				\hline 
				& P1 & P2 & P3 & P4 & P5 & P6 & P7 & P8 & P9 & P10 \\ 
				\hline \hline 
				Mean $(m/s^2)$ & 9,2686 & 9,6033 & 8,9072 & 9,8023 & 10,0307 & 9,6881 & 9,4592 & 9,7564 & 9,3649 & 9,7248 \\ 
				\hline 
				Standar Error & 0,1198 & 0,1138 & 0,1591 & 0,1632 & 0,0811 & 0,1469 & 0,1330 & 0,0829 & 0,1789 & 0,0761  \\ 
				\hline 
				Variance & 0,5451 & 0,4531 & 0,9619 & 0,9059 & 0,2303 & 0,7550 & 0,6723 & 0,2611 & 1,2465 & 0,2029  \\ 
				\hline 
				Minimum $(m/s^2)$  & 7,3327 & 7,8441 & 6,6046 & 7,3988 & 9,2361 & 7,4540 & 7,2124 & 8,7075 & 6,6458 & 8,8344  \\ 
				\hline 
				Maximum $(m/s^2)$  & 10,4333 & 11,0175 & 10,2034 & 11,4956 & 11,1127 & 10,9666 & 10,6110 & 10,7344 & 11,2452 & 10,6627 \\ 
				\hline 
				Count & 38 & 35 & 38 & 34 & 35 & 35 & 38 & 38 & 39 & 35 \\ 
				\hline 
				Level of Confidence 95\% & 0,2427 & 0,2312 & 0,3224 & 0,3321 & 0,1649 & 0,2984 & 0,2695 & 0,1679 & 0,3619 & 0,1547 \\ 
				\hline 
			\end{tabular} 
		}
		\caption{Tests comparison for experiment 8. Data from the smartphone.}
		\label{tabla:compromedio2}
	\end{center}
\end{table}

\begin{table}[htbp]
	\begin{center}
		\scalebox{1.1}{
		\begin{tabular}{|l|c|c|c|}
				\hline 
				& AC & SM & VID \\ 
				\hline \hline 
				Mean Value $(m/s^2)$  & 9,7216 & 9,5507 & 8,8730  \\ \hline 
				Standar Deviation (g) & 0,0268 & 0,0236 & 0,0390  \\ \hline 
				Minima $(m/s^2)$  & 8,1952 & 7,7075 & 10,9105   \\ 	\hline 
				Maximum $(m/s^2)$ & 10,9833 & 10,8392 & 6,0913  \\ \hline 
				Count & 116,6 & 36,5 & 29,5  \\	\hline 
				Level of Confidence 95\% & 0,1286 & 0,2554 & 0,4810  \\ 
				\hline 
			\end{tabular} }
		\caption{Comparison of the three methods (Acelerometer \textbf{Ac}, Smartphone \textbf{Sm} and Video Analysis \textbf{Vd}) for Experiment 8.}
		\label{tabla:compromedio4}
	\end{center}
\end{table}

The results for experiment 8, were similar to the other experiments. Here, it could observe that the mean values of the three methods appear to be close. However, there is not enough statistical evidence to conclude that the three methods are equivalents.

To develop a deeper analysis, it has been proposed a precision, variance and regression analysis.

\subsection{Precision Analysis}
To analyze the precision of each method and between them, it has been determined the parameter \textbf{quantity of information (QI)}  using the global parameters for each experiment with the three methods, see Table \ref{tabla:compromedio4}. The IQ is defined by the ratio between the amount of data $n$ and the variance or noise of the information $S^2$ as $(n/S^2)$.  The initial assumption is that the tests made with Ac are more precise, then verified with checking that the IQ higher. Thus,  it is possible to compare the IQ and stablish an estimation number, which provides the number of braking tests using Sm or Vd that it is necessary to achieve the same IQ than Ac.  The table \ref{tabla:Precision} shows the IQ and estimation numbers for the 8 experiments. For instance, the precision parameters obtained for experiment 8 presents that IQ for Ac is higher than the other methods, which means, the AC method is more precise. On another hand, the number of estimation for Sm is 4, namely, there are need 4 braking tests using Sm to get the same IQ as Ac. Besides, the number of the estimation of Vd for experiment 8 is 18, namely, there are need 18 braking tests using Vd to get the same IQ as Ac. The results present similar behavior for all experiments.

		\begin{table}[htbp]
		\begin{center}
			\scalebox{1}{
				\begin{tabular}{|c|c|c|c|c|c|}
					\hline 
					EXP	&	IQ-AC	&	IQ-SM	&	EN	&	IQ-VID	&	EN	\\ 	\hline \hline 
					2	&	713,8820	&	203,3801	&	4	&		&		\\ 	\hline 
					3	&	470,6407	&	209,2765	&	3	&	13,3863	&	36	\\ 	\hline 
					4	&	297,1109	&	192,7512	&	2	&		&		\\ 	\hline 
					5	&	348,4668	&		&		&	17,7927	&	20	\\ 	\hline 
					6	&	415,1366	&	49,2776	&	9	&	32,4358	&	13	\\ 	\hline 
					7	&	113,2596	&	30,2870	&	4	&	17,5972	&	7	\\ 	\hline 
					8	&	330,7808	&	82,8083	&	4	&	19,4280	&	18	\\ 	\hline
				\end{tabular} 
			}
			\caption{Precision Analisys. IQ corresponds to the Information Quantity  of each method and EN means the estimation number of Sm and Vd regarding to Ac.}
			\label{tabla:Precision}
		\end{center}
	\end{table}

The precision analysis permits to determine quantitatively that Ac is the most precise method using in this study. However, it is important to know the origin of variations between and within methods through variance analysis. The following procedure allowed a more detailed comparison, where it is possible to establish a hypothesis testing to determine if the mean values (deceleration) which have obtained with the different methods present or not statistical difference.

\subsection{Variance Analysis}
The variance analysis ANOVA studies the origin of variations between the methods, as well as within them. This technique examines the total variation of the experiments in components and comparing the mean values of each method.

The ANOVA considers the data acquired with each method ($k$ populations)  being independent and normally distributed. The approach established two hypotheses to test. The first one corresponds to the null hypothesis $H_0$ which considers the mean values ($ \mu_1, \mu_2, ... \mu_k $) of data for each method to be equal, namely they do not have significant statistical variations. On another hand, the ANOVA contemplates the alternative or research hypothesis $H_1$, which considers the values are not the same for all groups (methods).

If $y_{ij}$ is the $j$th observation of the $i$th method, $\bar{y_i}$ correponds to the mean of the all observations for each method and $\bar{y_i}$ is the mean value of the all $nk$ observations.
The total variability of $SS$ has been split in two terms. 
The variability between the methods $SS_m$ and within them $SM_e$. The last term is also known as error or aleatory variability.
%\begin{equation}
%\sum_{i=1}^{k} \sum_{j=1}^{n} ( y_{ij} - \bar{y})^2 = n\sum_{i=1}^{k} (\bar{y_i} - \bar{y})^2 + \sum_{i=1}^{k} \sum_{j=1}^{n} ( y_{ij} - \bar{y_i})^2
%\end{equation}

\begin{eqnarray}
 SS &= & \sum_{i=1}^{k} \sum_{j=1}^{n} ( y_{ij} - \bar{y})^2  \\
 SS_m & = &  n\sum_{i=1}^{k} (\bar{y_i} - \bar{y})^2 \\
 SS_e & = & + \sum_{i=1}^{k} \sum_{j=1}^{n} ( y_{ij} - \bar{y_i})^2 
\end{eqnarray}

For the ANOVA, the Fisher test has been made for testing the hypotheses (null and alternative) to study the means values similarities. The $F_{value}$ calculated from data permits to compare the variations or how far the data are scattered from the global mean, expressed as the ratio of the sum of the squares between groups divided by the sum of the square within a group, as described in Eq \ref{eq:Fvalue}.  The $F_{critical}$ was obtained, through the Fisher distribution adopting a level of significance of $\alpha=5\%$. In hypotheses testing, $\alpha$ is related to the probability to make an error type I, which means reject the null hypothesis given it is true. For the experiments, $F_{value}$ have been compared with $F_{critical}$. In ANOVA if  $F_{value}$ is larger than $F_{critical}$ The null hypothesis is rejected. 

\begin{figure}[h]
	\centering
	\includegraphics[width=0.7\linewidth]{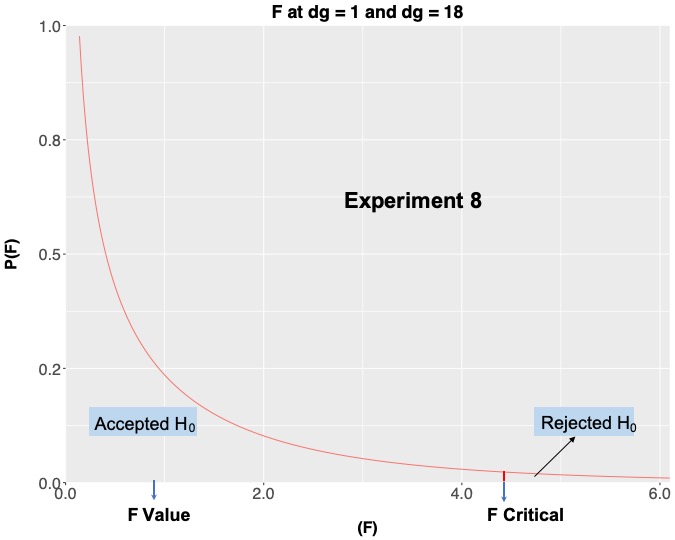}
	\caption{Fisher distribution for experiment 8 from Ac and Sm data. F$_{value}$ is lower than F$_{critical}$ Therefore, H$_{0}$ was accepted, which means that the expected values of Ac and Sm do not have significant statistical differences.}
	\label{fig:Fisher}
\end{figure}

For instance, Fig \ref{fig:Fisher} shows the F distribution for experiment 8 for data compared from Ac and Sm method. Here, the degree of freedom (df) of numerator was 1 and the (df) of denominator was 18, the F critical correspond to 4.41  and the F value to 1.96, thus, the null hypothesis has been accepted. As a consequence, the analysis for this experiment concludes that the mean values of the Ac and Sm methods do not have significant differences.

\begin{equation}
   F_{value}=   \frac{SS_m}{k-1}/ \frac{SS_e}{k(n-1)}
\label{eq:Fvalue}
\end{equation}

Table \ref{tabla:ANOVA2} summarizes variance analysis for the total sample obtained through the braking tests described in the previous section.  It exhibits $F_{value}$ and $F_{critical}$ to compare the mean values from  Ac, Sm, and Vd. For all experiments, the results present $F_{value}$  for Ac and Sm is lower than $F_{critical}$, which means that they do not have significant statistical variance. On the contrary, the   $F_{value}$ for Ac and Vd is bigger than $F_{critical}$, except for Experiment 5, which means their mean values present statistical variance, namely, there is evidence of differences between methods in the majority of experiments.

	\begin{table}[htbp]
	\begin{center}
	\scalebox{1}{
	\begin{tabular}{|c|c|c|c|c|c|c|c|c|c|c|c|}
		\hline 
		EXP & \multicolumn{2}{c|}{Ac-Sm}  & \multicolumn{2}{c|}{Ac-Vd}   \\ 

		\hline 
		 EXP 1	&	F$_{value}$	& F$_{critical}$	&	F$_{Value}$	&	 F$_{Critical}$		 \\ 	\hline
	
		EXP 2	&	0,001358 &  4,35 & N/A & N/A	 \\ 	\hline
		EXP 3	&	0,16392 & 4,49 & 5,0775 & 4,49	 \\ 	\hline
		EXP 4	& 0,01342	& 4,3 & N/A& N/A \\ 	\hline
		EXP 5	& N/A	& N/A & 3,4347&	4,41 \\ 	\hline
		EXP 6	&2,7495	& 4,3 & 24,2165 & 4,3	 \\ 	\hline
		EXP 7	&	0,02251& 4,3 & 30,4792 &	4,3	 \\ 	\hline
		EXP 8	&	1,9674 & 4,41 & 38,275 &	4,41	 \\ 	\hline
	
	\end{tabular} }
	\caption{ANOVA of Experiments}
	\label{tabla:ANOVA2}
	\end{center}
	\end{table}
	
\subsection{Regression Analysis}
To analyze the dependence relation between Sm and Ac, as well as Vd and Ac, it has used to simple linear regressions. The aim of this analysis has been visualized how the data acquired by the Sm and the Vd methods could predict the Ac method \cite{Dios10}. For each experiment, it assumed $Ac$ as the dependent variable and $Sm$ and $Vd$ as the independent variables. Letting $(x_{1}, y_{1}), (x_{2}, y_{2})...(x_{n}, y_{n})$ represent n data points, where n correspond to the number of tests for each experiment.  $x_{i}$ and $y_{i}$ are the mean values (mobile average) of the independent and dependent variables, respectively.  The simple linear regression can be described by two linear models, with and without intersection, as in Eq. \ref{RLT} and Eq. \ref{RL}, respectively.

\begin{eqnarray}
y_{i}& =  &\beta_0 + \beta_{1}x_{i}+\epsilon_{i}  \label{RLT} \\
y_{i}& = & \alpha x_{i}+\epsilon_{i} \label{RL}
\end{eqnarray}

In Eqs \ref{RLT}, \ref{RL} the errors $\epsilon_i$ are independently and normally distributed with a zero mean and a $\sigma^2$ variance. Because of the nature of the experiment, the model with intercept may not conceptually apply to the data, because a priori it is known that $x=0$ when $y=0$. However, it is not enough to justify regression through the origin. To determine which linear model may be a better fit, it has checked the models through some main regression parameters, such as $R^2$ and residual standard error $RSE$. $R^2$, described in Eq. \ref{R2} measures the proportion of variability in $y$ that can be explained using $x$. Whereas, $RSE$ given by Eq. \ref{RSE} provides an absolute measure of lack of fit of the model to the data

\begin{eqnarray}
R^2= \sum_{i} (y_{i}-\bar{y_{i}})- \sum_{i} (y_{i}-\hat{y_{i}})^{2}  /  \sum_{i} (y_{i}-\bar{y_{i}}) \label{R2} \\
RSE=\sqrt{\frac{1}{n-2} \sum_{i} (y_{i}-\hat{y_{i}})^{2}} \label{RSE}
\end{eqnarray}

\begin{figure}[h]
	\centering
	\includegraphics[width=1\linewidth]{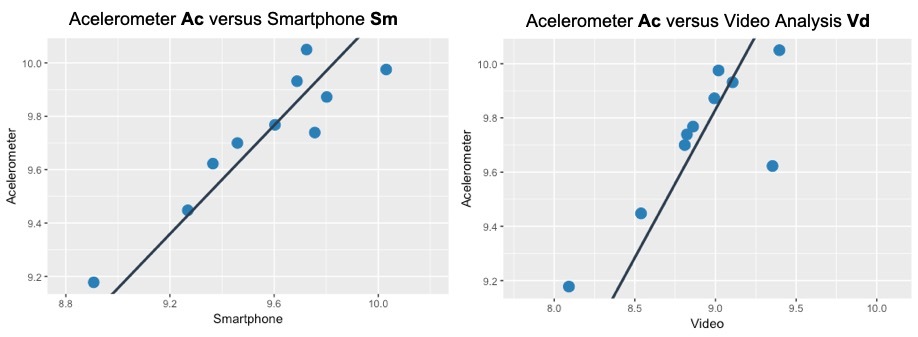}
	\caption{Shown linear regression models for experiment 8.  (a) Depicts the linear regresion for $Ac$ versus $Sm$. Here, the $R^2$ calculated was 0,9998 and $RSE$ was  0,1318. (b) Shown the linear regresion for $Ac$ versus $Vd$,  where $R^2$ obateined was 0,9994 and $RSE$ was 0,2521.  }
	\label{fig:RL-Exp8}
\end{figure}

For both models, which correspond to $Ac$ as a function of $Sm$ and $Ac$ versus $Vd$, depicted in Fig \ref{fig:RL-Exp8} (a) and Fig \ref{fig:RL-Exp8} (b), respectively. $R^2$ is close to 1, which means, the adjust is almost perfect. On another hand the $RSE$ is very small, verifying the good accuracy of the fitting. Therefore, this analysis presents that the mean values distribution obtained with $Ac$ could be estimated by $Sm$ and $Vd$ methods with very good accuracy. The linear regression analysis was made for all experiments, results showed that $R^2$ was always close to 1 and the $RSE$ close to 0 for both fitting models.

%Fig shows the linear regression analysis made for all experiments. The results present that $R^2$ was always close to 1 and the $RSE$ close to 0. The variance value $\sigma^2$ between the determination coefficient $R^2$ for all experiments was close to 0 for both fitting models, namely, all results pointing out to the same lineal relation, $Ac$ versus $Sm$ illustrated in Fig (a) and $Ac$ versus $Vd$ depicted in Fig (b). 

%
%Fig. shown the linear fitting for all experiments. Fig (a) $Ac$ versus $Sm$ and Fig (b) $Ac$ versus $Sm$.

%\begin{tabular}{c}
%
%      % after \\: \hline or \cline{col1-col2} \cline{col3-col4} ...
% \includegraphics [width=0.9\linewidth]{ImagenesVF/IMGexp8INT0.png} \\
%\includegraphics [width=0.9\linewidth]{ImagenesVF/IMGexp8vidINT0.png} 
%
%\end{tabular}
%
%	\begin{multline*}
%	P \left[(\bar{X_1} - \bar{X_2}) - t_{n_1 + n_2 - 2,1 - \dfrac{\alpha}{2}} \sqrt{\dfrac{(n_1 -1)S_1^2 + (n_2 - 1)S_2^2}{n_1 + n_2 - 2}} \sqrt{\dfrac{1}{n_1} + \dfrac{1}{n_2}}\right] \leq P(\mu_1 - \mu_2)\\
%	\leq P \left[ (\bar{X_1} - \bar{X_2}) t_{n_1 + n_2 - 2,1 - \dfrac{\alpha}{2}} \sqrt{\dfrac{(n_1 -1)S_1^2 + (n_2 - 1)S_2^2}{n_1 + n_2 - 2}} \sqrt{\dfrac{1}{n_1} + \dfrac{1}{n_2}}\right] = 1 - \alpha
%	\end{multline*}

\subsection{Confidence Intervals}
Considering previous results, the confidence intervals have been calculated to obtain a practical application of the use of method SM for braking test with vehicles in comparison to the use of Ac.
The $t$ student statistical was applied assuming a level of confidence of 95\% (or $\alpha = 5\%$), which means that at least 9 of 10 expected values for Ac fall within the intervals of confidence calculated from the data acquired with Sm and Vd. Table \ref{tabla:IC} shown, the upper limit $Up_{Lim}$ and the lower limit $Low_{Lim}$, belong to the confidence intervals, as well as the absolute error $\epsilon_{abs}$ and relative error $\epsilon_{rel}$, from the equations \ref{LL}, \ref{UL}, \ref{errorabs} and \ref{errorrel}, respectively.

	\begin{table}[htbp]
	\begin{center}
		\scalebox{1}{
			\begin{tabular}{|c|c|c|c|c|c|}
				\hline 
				Exp	&	$Low_{Lim} $	&	$Up_{Lim}$	&	$\epsilon_{abs}$	&	$\epsilon_{rel}$	\\ 	\hline \hline 
				EXP 2	&	-0,89	&	0,93	&	0,02	&	0,00		\\ 	\hline 
				EXP 3	&	-0,35	&	0,47	&	0,07	&	0,01	\\ 	\hline 
				EXP 4	&	-0,35	&	0,36	&	0,02	&  0,00		\\ 	\hline 
				EXP 6	&	-0,06	&	0,42	&	0,19	&	0,02		\\ 	\hline 
				EXP 7	&	-0,20	&	0,24	&	0,02	&	0,00		\\ 	\hline 
				EXP 8	&	-0,14	&	0,19	&	0,17	&	0,02		\\ 	\hline
			\end{tabular} 
		}
		\caption{The confidence intervals for the deceleration $(m/s^2)$ expected values with Ac from the data acquired with the Sm method.}
		\label{tabla:IC}
	\end{center}
\end{table}

\scriptsize{\begin{eqnarray}
(\bar{X_1} - \bar{X_2}) - t_{n_1 + n_2 - 2,1 - \frac{\alpha}{2}} \sqrt{\frac{(n_1 -1)S_1^2 + (n_2 - 1)S_2^2}{n_1 + n_2 - 2}} \sqrt{\frac{1}{n_1} + \frac{1}{n_2}} \label{LL} \\
((\bar{X_1} - \bar{X_2}) t_{n_1 + n_2 - 2,1  - \frac{\alpha}{2}} \sqrt{\frac{(n_1 -1)S_1^2 + (n_2 - 1)S_2^2}{n_1 + n_2 - 2}} \sqrt{\frac{1}{n_1} + \frac{1}{n_2}} \label{UL} 
\end{eqnarray}}

\normalsize

\begin{eqnarray}
\epsilon_{abs} & = & \Delta X_i =  | \bar{X} - X_i |  \label{errorabs} \\
\epsilon_{rel} & = & \frac{\Delta X_i }{\bar{X}}  \label{errorrel}
\end{eqnarray}

The intervals of confidence calculated above, permit apply the results of this study to a traffic accident reconstruction.  For instance, the deceleration means the value was 9,27 $m/s^2$ using Sm for tests in Exp 8. Considering the interval of confidence, it could be possible to assert that the mean value with Ac will be between 9,13 $m/s^2$ and 9,46 $m/s^2$. At last, it is possible to achieve the overall confidence interval from this study taking the lower and the upper value from all experiments.

%	\begin{eqnarray}
%	P \left[(\bar{X_1} - \bar{X_2}) - t_{n_1 + n_2 - 2,1 - \frac{\a
%lpha}{2}} \sqrt{\frac{(n_1 -1)S_1^2 + (n_2 - 1)S_2^2}{n_1 + n_2 - 2}} \\ 
%\sqrt{\frac{1}{n_1} + \dfrac{1}{n_2}}\right] \leq  \\ 
%P(\mu_1 - \mu_2) \leq \\
% P \left[ (\bar{X_1} - \bar{X_2}) t_{n_1 + n_2 - 2,1 - \frac{\alpha}{2}} \\
% \sqrt{\frac{(n_1 -1)S_1^2 + (n_2 - 1)S_2^2}{n_1 + n_2 - 2}} \sqrt{\frac{1}{n_1} + \frac{1}{n_2}}\right] = 1 - \alpha
%	\end{eqnarray}

\subsection{Friction Coefficient Estimation}

\begin{figure}[htbp]
	\centering
	\includegraphics[width=0.8\linewidth]{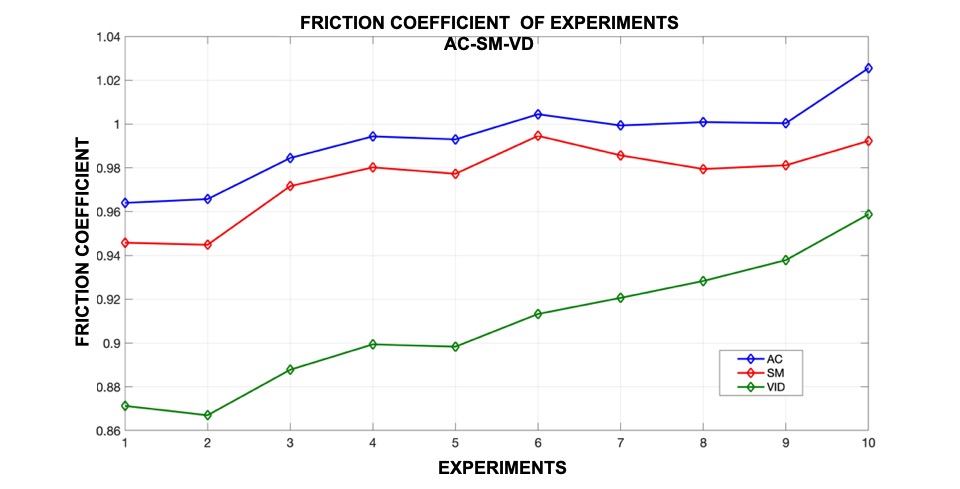}
	\caption{Friction coefficients in SZ obtained from the deceleration mobile average of experiments for each method Ac, Sm and Vd. }
	\label{fig:CF}
\end{figure}

To estimate the friction coefficients (FCs) , it is important to take into account the deceleration behavior during the braking test with the vehicle in Fig \ref{fig:grafic-p99}. As was mentioned above, there are two relevant zones the IZ, where the deceleration increase as a function of time and the SZ with fluctuating behavior of deceleration, which has been treated mathematically applying the moving average. In Fig \ref{fig:CF}, the three methods have been compared to estimate the friction coefficient in SZ, using the expected value of deceleration as a constant in that zone and applying the simplest physical model of uniformly accelerating motion and energy conservation, it could be asummed as a direct relation $\mu=a/g$  \cite{Lynn}. 
%
%\begin{figure}[htbp]
%	\centering
%	\includegraphics[width=1\linewidth]{ImagenesVF/CFACSMVID-1}
%	\caption{Video recording perpendicular to the vehicule trajectory}
%	\label{fig:CF}
%\end{figure}

Fig \ref{fig:CF} shows the friction coefficient calculated for SZ. It is important to consider that it reflects the different conditions of contexts of the experiments. Likewise, it is possible to observe that the absolute repeatability is difficult to achieve, despite the attempts to reproduce similar conditions, for instance, Experiment 6 and Experiment 7, those have the same context.

The FCs estimated with SM did not present important quantitative differences with the AC method. Whereas, the FCs obtained with video tests were a bit further away from AC. However, one of the main goals of obtaining FC is to know the initial velocity at the beginning of the braking maneuver. It was interesting to visualize that the FCs with three methods give as a result similar initial velocity values. Table \ref{tabla:Velocidad} exhibits the values of initial velocity in SZ $V_{\mu_{ZE}}$ for tests in Experiment 8. It is possible to observe that the average velocities calculated with FCs obtained with methods are close, with a difference no greater than $2$ Km/h. Although, is important to make more rigorous calculations including the propagation of the errors. 

 \begin{table}[htbp]
 	\begin{center}
 		\scalebox{0.7}{
 			\begin{tabular}{|c|c|c|c|c|c|c|c|c|c|c|c|}
 				\hline 
 				Variables & P1 & P2 & P3 & P4 & P5 & P6 & P7 & P8 & P9 & P10 & Prom.  \\ 
 		                  \hline

 				$V_{\mu_{ZE}} AC (Km/h)$		&	40,57	&	38,78	&	38,09	&	39,38	&	41,37	&	39,78	&	40,98	&	42,01	&	41,27	&	38,80 & 40,1 \\ 	\hline 
 				$V_{\mu_{ZE}} SM (Km/h)$	&	40,18	&	38,45	&	37,53	&	39,24	&	41,49	&	39,30	&	40,47	&	42,04	&	40,72	&	38,17 & 39,76 \\ 	\hline
 				$V_{\mu_{ZE}} VID (Km/h)$	&	38,56	&	36,94	&	35,77	&	37,58	&	39,34	&	38,10	&	39,05	&	39,98	&	40,69	&	37,52 & 38,35 \\ 	\hline
 			\end{tabular} 
 		}
 		\caption{Comparison of initial velocities obtained at the beginning of the SZ for Experiment 8, using the three methods Ac, Sm, and Vd. The $V_{\mu_{ZE}}$ was estimated with the friction coefficient average and distance of the SZ. }
 		\label{tabla:Velocidad}
 	\end{center}
 \end{table}

\section{Conclusion and Discussion}
The precision, variance and linear regression analysis of the deceleration data acquired during braking tests with a vehicle, stablished not significant differences between Ac and Sm. The variance study permitted to compare both methods through a fisher parameter. For all experiments, the $F_{Value}$ was lower than $F_{Critical}$, which represents a location into the probability distribution corresponding to the null hypothesis ($H_{0}$), namely, between the expected values of methods compared there are no statistical differences.

On another hand, the precision and variance analysis to compare the video processing $Vd$ and Accelerometer $Ac$ methods shown some differences between them. The $F_{Value}$ was not always lower than $F_{Critical}$  However, the linear regression depicted a very accuracy linear relation with an adjustment parameter almost 1, which means that the data from the $Vd$ could estimate directly the data from $Ac$ . 

The information quantity IC showed the importance of the number and the quality of data. The improvement of results may be possibly increasing the number of frames per second for video processing and the setting configuration of the smartphone to collect more data. 

This study opens the door for the feasible integration of the use of the easy access technology to be applied in the traffic accident reconstruction. This contribution will be very useful to low-middle income countries, where nowadays many experts are including parameters from foreign references into their calculations; among them the friction coefficient, which could not be appropriate to describe the real conditions to quantify the interaction road/vehicle in the traffic accident location.

\section{Acknowledgments}
The authors want to dedicate this paper to the late Professor Dr. Diosdado Baena; his guidance during the development of this project was of so much value that words are not sufficient to express how grateful that they are for his contributions. Thanks to Universidad Antonio Nari\~no especially to VCTI (Vicerrector\'ia de Ciencia, Tecnologo\'ia e Innovaci\'on UAN) for financial support and for the trust you have placed on the Research Line of Forensic Physics of Faculty of Science. The authors want also to acknowledges To Policia Nacional de Colombia particularly to ESEVI (Escuela de Seguridad Vial) and ESGON (Escuela Gabriel Gonzales) for providing the spaces for braking tests and the constant support.

\appendix
    \section{Protocol to the simultaneous use of Smartphone, Video Recording and Accelerometer to measure deceleration during an emergency braking test with vehicles}

This study proposes a procedure for the simultaneous use of three methods to measure deceleration during an emergency braking maneuver with vehicles.  This protocol integrates the following standards:
\newline
ISO21994 for Stopping distance at straight-line braking with ABS -Open-loop test method \cite{ISO21994}.
\newline
Measurement of Vehicle-Roadway Frictional Drag SAEJ2505 \cite{SAE2505}.
\newline
Measurement of the deceleration factor in vehicle brakes with the smartphone by CEIRAT.
\newline
To prepare experimental test it must take into account:
\begin{itemize}

\item Security.
\item Status and operation of active and passive vehicle safety systems to be used.
\item Road and environmental conditions.
\item Checking of measure equipment.
\end{itemize}

To follow, according to standards mentioned, it has been summarized the relevant parameters and actions, which it must take into account before the braking test:

\begin{itemize}

\item To guarantee the test conditions, all of them should be made in the same road section.
\item  It is necessary to guarantee that neither wear of the tread no frequently braking modifies the road conditions.
\item  Apply the procedure of conditioned on tires and brakes on the test track before experiments.
\item The initial condiction of driving should be on  a straight section of road.
\item The brakes then should be rapidly actuated and
continued with sufficient force to maintain wheel lock and/or ABS operation until de vehicle stops. Minima force must be 500 N.
\item During braking do not apply important steering corrections .
\item For one experiment, namely, a sequence of measuring, it should be made  10 valid  measures.

\end{itemize}

The simultaneous use of measuring devices  follows the recommendations

\subsection{Measuring of deceleration using smartphone:}

\begin{itemize}
\item Before starting the braking tests, the driver must take time to become acquainted with the operation of the vehicle.
\item Before the beginning of testing, the co-driver should place the calibrated smartphone into the horizontal plane of the vehicle, subject this to an object as close as possible to the center of mass of the vehicle.
\item Before initializing the vehicle motion, the co-driver must start the capture of the data with the smartphone app. Before starting the braking the driver should maintain the speed selected constant for 1,5 sec.
\item Once the vehicle has been stopped, the co-driver must stop the application and record de data obtained in the smartphone.
\item Once braking test has finished, the co-driver has to save or download the data set of test (format csv or xls).
\end{itemize}

\subsection{Measuring of deceleration using video recording}

\begin{figure}[htbp]
	\centering
	\includegraphics[width=0.8\linewidth]{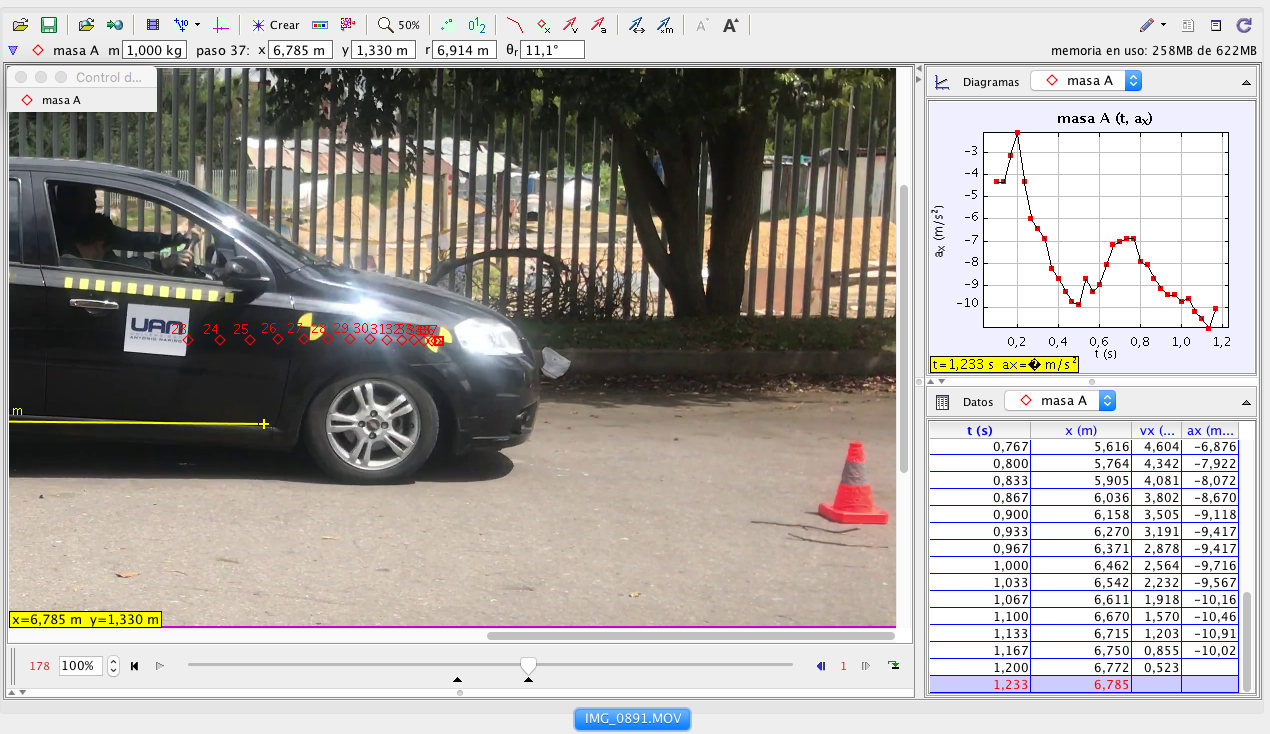}
	\caption{Video recording must be perpendicular to  the vehicle trajectory to extract the kinematic parameters of braking test with Tracker (free software).}
	\label{fig:tracker5}
\end{figure}

\begin{itemize}

\item For recording video it is important to have a tripod that allows the positioning of camera perpendicular to the via and trajectory of the vehicle during deceleration, being careful to avoid movements of recording device.
\item Two cones should be located to indicate the start and end of the braking maneuver as shown in Fig \ref{fig:tracker5}  The cones will be the reference of the film close up.
\item Each video recording is synchronized with the start of each test. The number assigned to the test should be the same for each device used simultaneously.
\item Videos recordings are analysis with free software Tracker. It is recommended to check guideline \cite{tracker}.
\end{itemize}

\subsection{Measuring of deceleration using accelerometer VC400PC}

\begin{itemize}
\item The initial speed for the test should be programmed with the accelerometer in mode braking. When the vehicle reached the speed of work, brakes should be activated with beeps of the accelerometer.
\item Once the vehicle stop, the accelerometer has registered data for time, distance, deceleration, and speed. For operating the device see guideline \cite{VC4000}.
\end{itemize}

\bibliography{ArxivRAT}

\end{document}